\documentclass[12pt]{article}

\usepackage[a4paper,margin=1in]{geometry}  
\usepackage{graphicx}
\usepackage{float}
\usepackage{booktabs}
\usepackage{amsmath,amssymb}
\usepackage{authblk}
\usepackage{setspace}
\usepackage{titlesec}
\usepackage{abstract}
\usepackage{url}
\usepackage{braket}
\usepackage{array}
\usepackage{hyperref}
\usepackage{xcolor}

\titleformat{\section}{\normalfont\large\bfseries}{\thesection.}{1em}{}
\titleformat{\subsection}{\normalfont\normalsize\bfseries}{\thesubsection.}{1em}{}

\title{\textbf{Hemispheric-Specific Coupling Improves Modeling of Functional Connectivity Using Wilson--Cowan Dynamics}}

\author[1]{Ramiro Pl\"{u}ss}
\author[2]{Hern\'an Villota}
\author[3,4]{Patricio Orio}

\affil[1]{Instituto Tecnológico de Buenos Aires (ITBA), CABA, Argentina}
\affil[2]{ Institución Universitaria de Envigado (IUE), Envigado, Colombia}
\affil[3]{Insituto de Neurociencias, Facultad de Ciencias, Universidad de Valparaíso, Valparaíso, Chile}
\affil[4]{Centro Interdisciplinario de Neurociencia de Valparaíso, Valparaíso, Chile}

\date{}

\begin{document}
\maketitle

\begin{abstract}
Large-scale neural mass models have been widely used to simulate resting-state brain activity from structural connectivity. In this work, we extend a well-established Wilson--Cowan framework by introducing a novel hemispheric-specific coupling scheme that differentiates between intra-hemispheric and inter-hemispheric structural interactions. We apply this model to empirical cortical connectomes and resting-state fMRI data from matched control and schizophrenia groups. Simulated functional connectivity is computed from the band-limited envelope correlations of regional excitatory activity and compared against empirical functional connectivity matrices. Our results show that incorporating hemispheric asymmetries enhances the correlation between simulated and empirical functional connectivity, highlighting the importance of anatomically-informed coupling strategies in improving the biological realism of large-scale brain network models.
\end{abstract}

\textbf{Keywords:} Wilson--Cowan model, structural connectivity, functional connectivity, schizophrenia, neural dynamics, hemispheric coupling

\section{Introduction}

Schizophrenia is a severe neuropsychiatric disorder characterized by profound disruptions in thought processes, perception, and affect regulation. Mounting evidence indicates that these symptoms are associated with altered functional connectivity (FC) across large-scale brain networks, particularly within and between the default mode, frontoparietal control, and salience networks~\cite{griffa2015, vohryzek2020}. Understanding how these functional abnormalities arise from the underlying anatomical organization, or structural connectivity (SC), remains a central question in computational psychiatry. Structural connectomes derived from diffusion MRI provide a static scaffold upon which neural dynamics unfold. However, the mapping between SC and FC is complex due to nonlinear interactions and modulatory processes that shape brain activity. Computational models of whole-brain dynamics offer a principled framework to investigate this SC--FC relationship. In particular, the Wilson--Cowan neural mass model provides a biologically grounded formalism to simulate excitatory and inhibitory population dynamics, enabling the generation of realistic large-scale activity patterns~\cite{wilson1972, wilson1973}. In this work, we simulate FC from empirical cortical SC matrices obtained from matched control and schizophrenia groups. To isolate cortical dynamics and minimize confounds related to deep nuclei, we restrict our analysis to cortical regions. We first implement a classic whole-brain model with uniform global coupling strength ($G$), and then introduce a novel hemispheric-specific coupling strategy that distinguishes intra- ($G_1$) and inter-hemispheric ($G_2$) connectivity. This extension is motivated by anatomical and functional asymmetries previously reported in both healthy and clinical populations~\cite{griffa2015}. We evaluate the capacity of each model variant to reproduce empirical resting-state FC, using Pearson correlation as a measure of SC--FC correspondence. Additionally, we analyze group differences and explore the role of hemispheric specificity in capturing altered integration patterns observed in schizophrenia. Our results show that introducing hemispheric coupling parameters improves the model's performance and may help reveal biologically meaningful asymmetries in large-scale brain dynamics.

\section{Material and Methods}
\subsection{Empirical Dataset and Brain Network Estimation}

We employed a publicly available dataset containing diffusion and functional MRI scans from 27 patients with schizophrenia and 27 demographically matched healthy controls~\cite{vohryzek2020, gutierrez2020}. These data were acquired using a 3T Siemens Trio scanner and include high-resolution T1-weighted images, diffusion spectrum imaging (DSI), and resting-state fMRI. Participants in the patient group met DSM-IV criteria for schizophrenia or schizoaffective disorder, and were recruited from the Lausanne University Hospital. Brain network construction was performed using the Connectome Mapping Toolkit~\cite{daducci2012}. Cortical segmentation was based on the Lausanne 2008 multi-scale parcellation scheme~\cite{cammoun2012}, from which we selected the 83-region resolution to balance anatomical specificity with computational feasibility. SC matrices were derived from deterministic streamline tractography applied to the DSI data, with edge weights defined as normalized streamline counts accounting for surface area and length biases~\cite{hagmann2008}. FC was estimated as the absolute Pearson correlation between regional mean BOLD signals after standard preprocessing, including motion correction, nuisance regression, spatial smoothing, and band-pass filtering~\cite{griffa2017, smith2009}. This framework provides subject-level structural and functional networks with matched parcellations, enabling direct comparison and simulation of brain dynamics constrained by individual connectomes. To reduce inter-subject variability and highlight group-level trends, we computed group-average SC and FC matrices separately for the control (CTRL) and schizophrenia (SCHZ) cohorts.Since all individual SC and FC matrices are symmetric by construction, owing to the undirected nature of tractography and the use of Pearson correlation, the resulting group-average matrices are also symmetric. This approach allows us to focus on global network properties and systematic differences between groups, rather than individual-specific fluctuations. By simulating neural dynamics on the group-averaged SC using the Wilson--Cowan model, we can directly assess how alterations in anatomical connectivity influence the emergence of functional interactions at the population level. Furthermore, the use of averaged matrices facilitates a controlled exploration of model parameters, such as global versus hemispheric-specific coupling schemes, enabling us to investigate how different modes of interaction (intra-hemispheric vs. inter-hemispheric connectivity) affect the model’s ability to reproduce empirical FC. This is particularly relevant in schizophrenia, where disruptions in inter-hemispheric integration have been consistently reported. Finally, from a translational perspective, the group-level SC and FC matrices serve as clinical reference profiles, against which deviations at the individual level could be later evaluated. This provides a meaningful scaffold for assessing the impact of pathological alterations and testing hypotheses about network-level mechanisms underlying psychiatric disorders. The 83-region resolution includes both cortical and subcortical structures. To restrict the analysis to cortical dynamics, we excluded subcortical regions and the brainstem, resulting in 68 cortical ROIs (34 per hemisphere). This reduction ensures compatibility with the Wilson--Cowan neural mass model, which is defined at the mesoscopic cortical level. All modeling and simulation steps were performed on these $68 \times 68$ group-average matrices.

\subsection{Wilson--Cowan Network Model}

We employed a whole-brain neural mass model based on the Wilson--Cowan equations~\cite{wilson1972, wilson1973}, which describe the dynamics of coupled excitatory ($E$) and inhibitory ($I$) neural populations. Each cortical region $k$ is modeled as a pair of interacting populations governed by Eq.~\eqref{eq:wcE} and Eq.~\eqref{eq:wcI}.

\begin{align}
\tau_e \frac{dE_k}{dt} &= -E_k + (1 - r_e E_k) S_e\left(a_{ee} E_k - a_{ei,k} I_k + G \sum_{\substack{l=1 \\ l \neq k}}^{N} C_{kl} E_l + P \right) \label{eq:wcE} \\
\tau_i \frac{dI_k}{dt} &= -I_k + (1 - r_i I_k) S_i\left(a_{ie} E_k - a_{ii} I_k + Q \right) \label{eq:wcI}
\end{align}

Here, \( \tau_e \) and \( \tau_i \) are the time constants for excitatory and inhibitory populations, respectively; \( r_e \) and \( r_i \) modulate population refractoriness; and \( a_{\alpha\beta} \) are synaptic weight parameters. The matrix \( C_{kl} \) encodes the normalized structural connectivity between regions, and \( G \) is a global coupling parameter. \( P \) and \( Q \) represent constant external inputs. The transfer function \( S(x; \mu, \sigma) \) is a sigmoidal nonlinearity that governs the population response, where \( \mu \) defines the activation threshold and \( \sigma \) controls the slope of the transition. Its general form is given by Eq.~\eqref{eq:sigmoid}.

\begin{equation}
S(x) = \frac{1}{1 + \exp\left( -\frac{x - \mu}{\sigma} \right)} \label{eq:sigmoid}
\end{equation}

This formulation has been widely used to simulate large-scale brain dynamics and to investigate the emergence of functional connectivity from anatomical structure~\cite{deco2013}. To derive simulated FC from the model, we extract the excitatory activity \( E_k(t) \) from each region after a transient period and compute its analytic signal using the Hilbert transform. The magnitude of this signal provides the amplitude envelope, which captures the slow oscillatory modulation of neural activity. Simulated FC is then defined as the Pearson correlation between the amplitude envelopes of all pairs of regions. This method follows standard procedures for extracting resting-state FC from band-limited neural signals, and allows for direct comparison with empirical FC matrices derived from resting-state fMRI. Prior to applying the Hilbert transform, the signals are band-pass filtered in the 0.01--0.1~Hz range to match the frequency content of resting-state BOLD fluctuations.

\subsection{Inhibitory Synaptic Plasticity}

To enhance the model’s stability and biological plausibility, we adopt a inhibitory synaptic plasticity (ISP) mechanism~\cite{abeysuriya2018}. This rule dynamically modulates the inhibitory-to-excitatory coupling $a_{ei}(t)$ within each cortical region based on local activity levels. The weight evolves according to Eq.~\eqref{eq:plasticity}.

\begin{equation}
\tau_p \frac{d a_{ei}}{dt} = I(E - \rho_E)
\label{eq:plasticity}
\end{equation}

The parameter $\tau_p$ is the adaptation time constant and $\rho_E$ is a target excitatory level. This form of homeostatic ISP promotes balanced excitation-inhibition interactions, preventing runaway excitation and enabling sustained oscillatory activity across a wide range of coupling regimes. By integrating Eq.~\eqref{eq:plasticity} alongside the Wilson--Cowan dynamics, the model remains compatible with empirical resting-state activity, while improving robustness and interpretability. Similar ISP rules have been shown to support biologically realistic network dynamics in large-scale simulations~\cite{abeysuriya2018, deco2013}.

\subsection{Hemispheric-Specific Coupling}

\begin{figure}[h]
    \centering
    \includegraphics[width=0.75\textwidth]{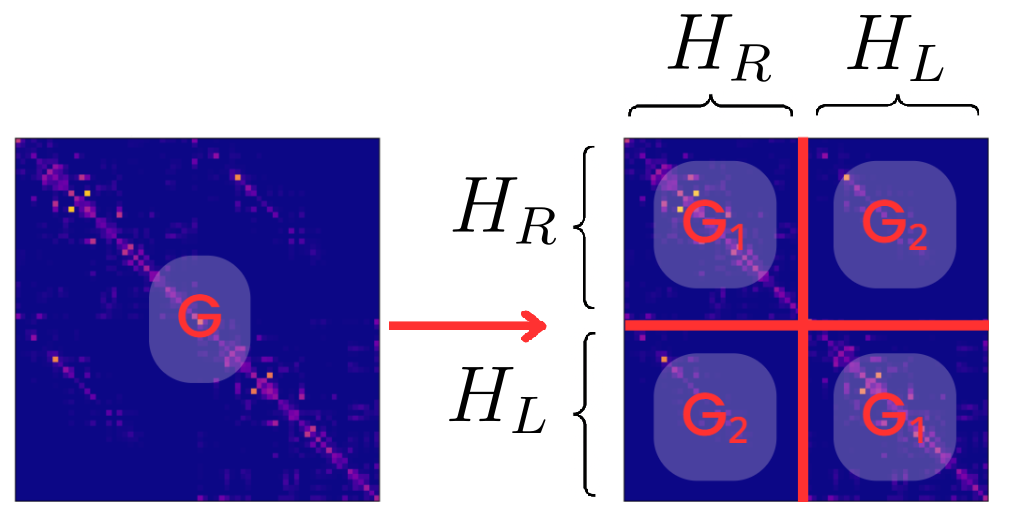}
    \caption{
    Coupling schemes in the Wilson--Cowan model. The left panel illustrates global coupling using a single parameter \( G \), which uniformly modulates connectivity across all regions. The right panel shows hemispheric-specific coupling, where \( G_1 \) modulates intra-hemispheric connections and \( G_2 \) modulates inter-hemispheric ones. The structural matrix is reordered such that the first 34 regions correspond to the right hemisphere (\( H_R \)) and the last 34 to the left hemisphere (\( H_L \)), resulting in a quadrant structure that reflects the anatomical division.
    }
    \label{fig:coupling_schemes}
\end{figure}

The standard formulation uses a single global parameter \( G \) to scale all long-range excitatory inputs. However, anatomical studies have shown that inter-hemispheric connections are sparser and more variable than intra-hemispheric ones, and that this asymmetry may be exaggerated in schizophrenia~\cite{griffa2015}. We introduce a modified coupling scheme that differentiates between intra- and inter-hemispheric projections via two parameters, \( G_1 \) and \( G_2 \), respectively. Fig.~(\ref{fig:coupling_schemes}) illustrates the difference between the global and hemispheric-specific coupling schemes applied to the structural connectivity matrix.

In the global case, all regions are influenced uniformly. In the hemispheric-specific scheme, the matrix is divided into four blocks: intra-hemispheric connections within \( H_R \) and \( H_L \) are scaled by \( G_1 \), while inter-hemispheric connections (between \( H_R \) and \( H_L \)) are scaled by \( G_2 \). This block-wise parameterization allows the model to capture asymmetric integration patterns across hemispheres. Specifically, the long-range term in Eq.~\eqref{eq:wcE} is replaced by Eq.~\eqref{eq:G1G2}.\\

\begin{equation}
G \sum_{\substack{l=1 \\ l \neq k}}^{N} C_{kl} E_l \quad \longrightarrow \quad \sum_{\substack{l=1 \\ l \neq k}}^{N} \left[ G_1 \delta_{kl}^{\text{intra}} + G_2 \delta_{kl}^{\text{inter}} \right] C_{kl} E_l, \label{eq:G1G2}
\end{equation}

where \( \delta_{kl}^{\text{intra}} = 1 \) if regions \( k \) and \( l \) belong to the same hemisphere, and \( \delta_{kl}^{\text{inter}} = 1 \) otherwise. This extension allows the model to explore the hypothesis that intra-hemispheric and inter-hemispheric coupling are differentially disrupted in schizophrenia. The parameters \( G_1 \) and \( G_2 \) were systematically varied to assess their impact on the correlation between simulated and empirical functional connectivity, as well as their influence on functional network topology.

\subsection{Parameter values}

All simulations were based on the Wilson--Cowan model. The transfer function \( S(x) \), defined in Eq.~\eqref{eq:sigmoid}, was parameterized with \( \mu = 1.0 \) and \( \sigma = 0.25 \), modulating the population response gain. The model included activity-dependent adaptation on the inhibitory-to-excitatory synaptic weight, initialized as \( a_{ei}(0) = 2.5 \), and evolving according to the excitatory-inhibitory interaction. Other synaptic weights were fixed: \( a_{ee} = 3.5 \), \( a_{ie} = 3.75 \), and \( a_{ii} = 0 \). Refractoriness coefficients were set to \( r_e = r_i = 0.5 \), with a threshold adaptation term \( \rho_E = 0.14 \) governing plasticity. Time constants were set to \( \tau_e = 10\,\mathrm{ms} \), \( \tau_i = 20\,\mathrm{ms} \), and \( \tau_{\mathrm{p}} = 1 \mathrm{s} \) for synaptic adaptation. The simulation was integrated using an Euler method with time step $\Delta t_{\mathrm{sim}} = 1 \times 10^{-4}\,\mathrm{s}
$, downsampled to $\Delta t = 1 \times 10^{-3}\,\mathrm{s}
$. A transient period of 100 seconds was discarded before collecting 100 seconds of simulated activity. Gaussian noise was introduced in the excitatory input with standard deviation \( \sqrt{D/\Delta t_{\mathrm{sim}}} \), where \( D = 2 \times 10^{-3} \). Empirical structural and functional connectivity matrices, normalized and restricted to 68 cortical regions, were used to couple regional dynamics and evaluate model performance. Each simulation was initialized with \( E_0 = I_0 = 0.1 \), and unless otherwise specified, a single realization per parameter pair was executed. We performed grid searches over coupling parameters in both models: $G$ in the global coupling scheme, and $G_1$, $G_2$ in the hemispheric-specific variant. In all cases, parameters were explored over the range \( G \in [0.1, 30.0] \). In the global model, $G$ uniformly scales all long-range excitatory inputs. In the hemispheric-specific model, by contrast, $G_1$ modulates intra-hemispheric coupling, whereas $G_2$ controls inter-hemispheric connectivity. Model performance was assessed by comparing the simulated functional connectivity with empirical FC using Pearson correlation and the Euclidean distance between the upper-triangular entries of the respective matrices.

\section{Results}

We analyze the empirical connectivity matrices for the CTRL and SCHZ groups to highlight their main differences and evaluate how well the simulations reproduce the observed FC patterns. We systematically assess the performance of a whole-brain model based on Wilson--Cowan dynamics. Our analysis is structured around two modeling approaches. First, we consider a single global coupling parameter, $G$, which uniformly scales the influence of all structural connections across the network. Second, we introduce two distinct coupling parameters: $G_1$ for intra-hemispheric and $G_2$ for inter-hemispheric interactions. This parametrization allows us to explore whether differentiating coupling within and between hemispheres improves the model's ability to replicate empirical patterns, particularly in light of known hemispheric asymmetries in schizophrenia. For each configuration, we perform parameter sweeps and compute both the pearson correlation and root mean square error between the simulated and empirical FC matrices. These metrics quantify the model’s accuracy and guide the selection of optimal parameters for each group.

\subsection{Topological Analysis of Empirical Structural Connectivity}

\begin{figure}[h]
    \centering
    \includegraphics[width=0.95\textwidth]{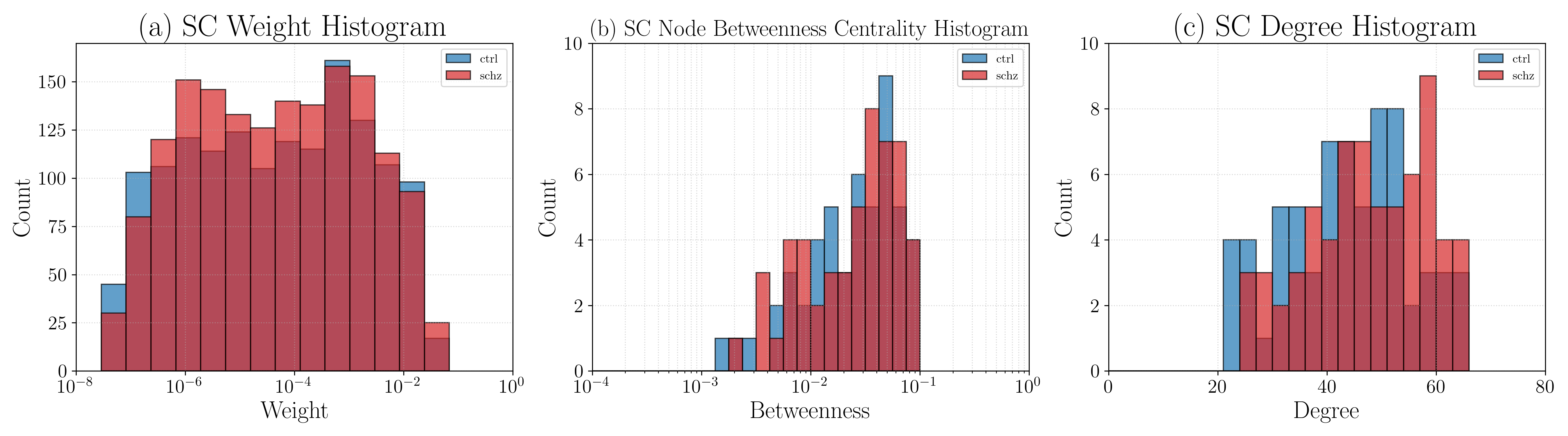}
    \caption{
        Comparison of SC weighted network properties between the CTRL and SCHZ groups, based on empirical group-average matrices. Histogram of 
            \textbf{(a)} connection weights, 
            \textbf{(b)} node betweenness centrality and
            \textbf{(c)} degree distribution.
            All metrics were computed from the empirical SC matrices restricted to cortical regions.}
    \label{fig:sc_distributions_ctrl_schz}
\end{figure}

We computed a set of global network metrics on the average SC matrices of the CTRL and SCHZ groups, including clustering coefficient, modularity, average shortest path length, global efficiency, mean strength, and the small-world index. Although minor numerical differences were observed between groups, no consistent or significant trends emerged that would support robust topological distinctions at the global scale. Then we compare SC weighted network histograms between the two groups, as shown in Fig.~\ref{fig:sc_distributions_ctrl_schz}.

Fig.~(\ref{fig:sc_distributions_ctrl_schz}a) displays the histogram of SC weights based on group-average matrices. Both groups exhibit a predominance of weak connections, consistent with the known sparsity of long-range anatomical projections. However, the SCHZ group shows a slightly higher count of weaker edges, suggesting a reduction in overall anatomical connectivity strength compared to the CTRL group. In Fig.~(\ref{fig:sc_distributions_ctrl_schz}b), the histogram of node betweenness centrality reveals that the CTRL group tends to distribute intermediary structural roles across a broader range of nodes, as indicated by a slightly wider spread in centrality values. In contrast, the SCHZ group exhibits a more concentrated profile, with fewer nodes contributing to long-range communication pathways. Nonetheless, the overall difference between the two groups remains modest at the level of group-average SC. In Fig.~(\ref{fig:sc_distributions_ctrl_schz}c), CTRL and SCHZ degree distributions are largely similar, with only subtle shifts, though the SCHZ group appears to exhibit fewer high-degree hubs.

\subsection{Wilson--Cowan Model with Uniform Global Coupling \( G \)}

To investigate the relationship between structural and functional connectivity (SC–FC) in both CTRL and SCHZ groups, we simulated the Wilson–Cowan model on empirical structural connectomes using a single global coupling strength parameter \( G \). We evaluated a range of \( G \) values logarithmically distributed from 0.1 to 30.0. For each group and each value of \( G \), we generated $10$ simulations, computed the resulting FC matrices, and compared them to the empirical FC using two complementary metrics: the Pearson correlation $(r)$ and the root mean squared error (RMSE).\\

\begin{figure}[H]
    \centering
    \includegraphics[width=0.95\textwidth]{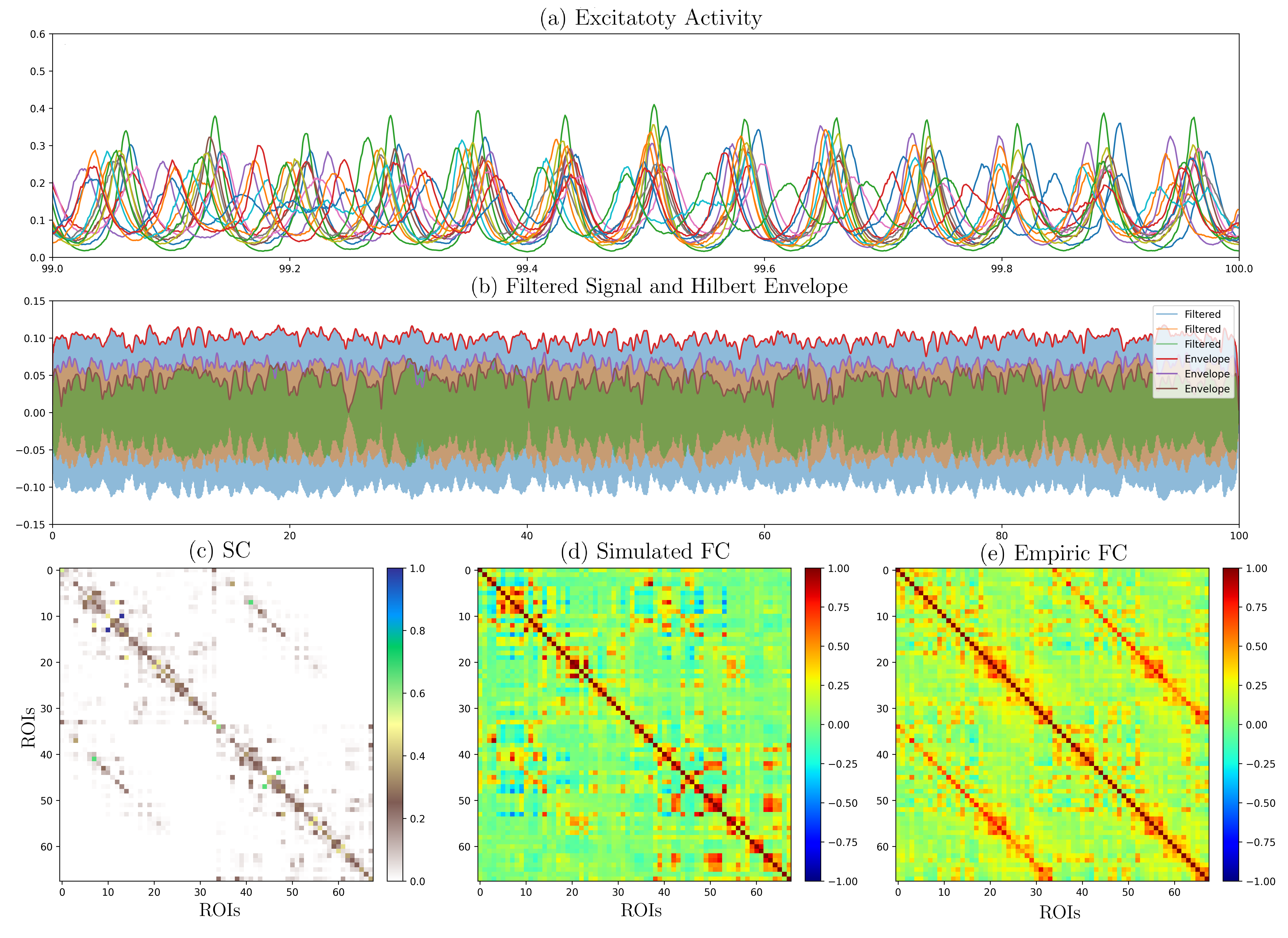}
    \caption{Group-averaged best-performing simulation for the SCHZ group using global coupling \( G = 1.0 \). \textbf{(a)} Excitatory activity, \textbf{(b)} Hilbert envelope, \textbf{(c)} SC, \textbf{(d)} Simulated FC \textbf{(e)} and empirical FC. The simulation yielded $ r = 0.37$, RMSE $= 0.22$.}
    \label{fig:bestG_schz}
\end{figure}

To illustrate the best-performing simulation under the global coupling model, Fig.~(\ref{fig:bestG_schz}) presents the results for the SCHZ group with \( G = 1.0 \). This instance yielded \( r = 0.37 \) and  \(\text{RMSE} = 0.22 \) with respect to the empirical FC. The top row shows the excitatory activity traces in Fig.~(\ref{fig:bestG_schz}a) and the filtered BOLD envelope in Fig.~(\ref{fig:bestG_schz}b). The bottom row displays the structural connectivity matrix in Fig.~(\ref{fig:bestG_schz}c), the simulated functional connectivity in Fig.~(\ref{fig:bestG_schz}d), and the empirical FC in Fig.~(\ref{fig:bestG_schz}e). The simulated FC captures the overall block organization and inter-hemispheric symmetry observed in the empirical pattern. As observed in Fig.~(\ref{fig:bestG_schz}d), a noticeable difference appears in the off-diagonal blocks, which are associated with inter-hemispheric connections in the network. The best-performing case for the CTRL group yielded \( r = 0.32 \) and \(\text{RMSE} = 0.24 \), indicating a slightly worse fit.\\

\begin{figure}[H]
    \centering
    \includegraphics[width=0.95\textwidth]{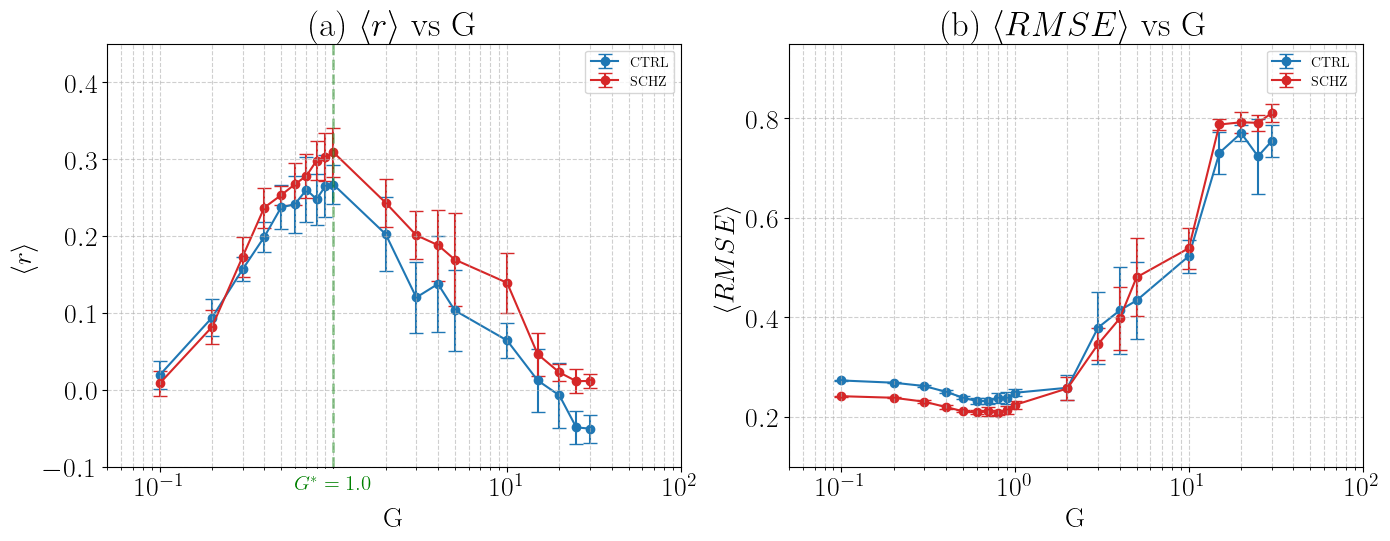}
    \caption{\textbf{(a)} Average pearson correlation $(\braket{r})$ and \textbf{(b)} average RMSE $(\braket{RMSE})$ between simulated and empirical FC as a function of global coupling strength \( G \), for both CTRL and SCHZ groups. Error bars indicate ±1 standard deviation across 10 simulations per value of \( G \).}
    \label{fig:avg_corr_rmse_vs_G_scale_1}
\end{figure}

Fig.~(\ref{fig:avg_corr_rmse_vs_G_scale_1}) shows the average Pearson correlation (\( \braket{r} \)) and the average root mean square error (\( \braket{\text{RMSE}} \)) across simulations as a function of the global coupling parameter \( G \), along with one standard deviation. The CTRL group reached a peak \( \braket{r} = 0.27 \pm 0.03 \), while the SCHZ group achieved a higher maximum of \( \braket{r} = 0.30 \pm 0.03 \), both groups at \( G = 1.0 \).\\

In both cases, the RMSE curve reached its minimum near the same values of \( G \), reinforcing the identification of optimal coupling regimes. Specifically, the lowest RMSE values were \( \braket{\text{RMSE}} = 0.25 \pm 0.01 \) for the CTRL group and \( \braket{\text{RMSE}} = 0.20 \pm 0.01 \) for the SCHZ group, both groups occurring at \( G = 1.0 \). On average, the SCHZ group achieved higher correlation values across simulations, suggesting that the model reproduces functional connectivity more accurately in patients than in controls under this global coupling scheme.\\

\subsection{Wilson--Cowan Model with Distinct Coupling Parameters \( G_1 \) and \( G_2 \)}

\begin{figure}[H]
    \centering
    \includegraphics[width=1.05\textwidth]{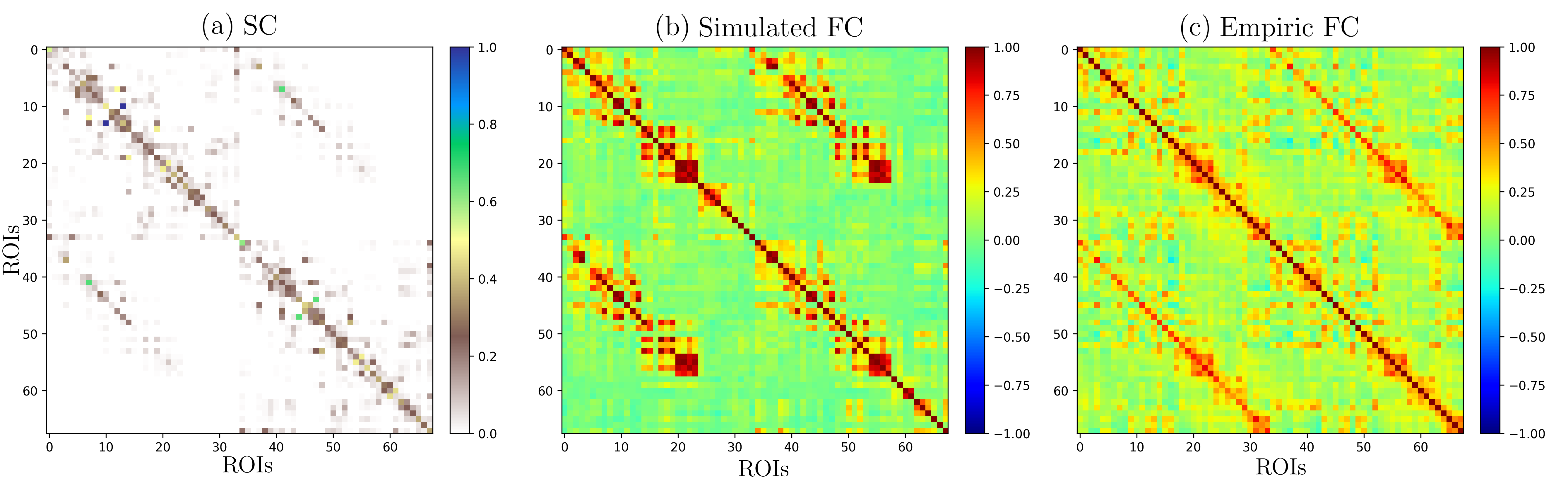}
    \caption{Group-averaged best-performing simulation for the SCHZ group using hemispheric-specific coupling parameters \( G_1 = 0.80 \), \( G_2 = 15.00 \). \textbf{(a)} Structural Connectivity (SC) matrix, \textbf{(b)} Simulated Functional Connectivity (FC), and \textbf{(c)} Empirical FC. The simulation yielded a correlation of \( r = 0.47 \) and RMSE = 0.20. Panels (a) and (c) correspond to the same subject as in Fig.~(\ref{fig:bestG_schz}c) and Fig.~(\ref{fig:bestG_schz}e), respectively.}
    \label{fig:bestG1G2_schz}
\end{figure}

To evaluate whether distinguishing intra-hemispheric and inter-hemispheric interactions improves model performance, we extended the global coupling scheme by introducing two separate parameters: \( G_1 \) for intra-hemispheric and \( G_2 \) for inter-hemispheric connectivity. To illustrate the best-performing simulation under the hemispheric-specific coupling model, Fig.~(\ref{fig:bestG1G2_schz}) presents the results for the SCHZ group with \( G_1 = 0.80 \) and \( G_2 = 15.00 \).

This configuration yielded a correlation of \( r = 0.47 \) and \(\text{RMSE} = 0.20\) when compared to the empirical FC. Fig.~(\ref{fig:bestG1G2_schz}) includes the structural connectivity matrix used for the simulation in Fig.~(\ref{fig:bestG1G2_schz}a), the resulting simulated FC in Fig.~(\ref{fig:bestG1G2_schz}b), and the empirical FC used for comparison in Fig.~(\ref{fig:bestG1G2_schz}c). The simulated FC replicates the modular structure and bilateral symmetry of the empirical matrix, with notable improvements in inter-hemispheric correlations. As observed in Fig.~(\ref{fig:bestG1G2_schz}b), the enhanced fit is particularly evident in the preservation of off-diagonal block structures, reflecting effective modeling of cross-hemispheric interactions. The best-performing case for the CTRL group yielded \( r = 0.40 \) and \(\text{RMSE} = 0.20 \) at $G_1 = 1.0$ and $G_2 = 15.0$, indicating a slightly worse fit. On average, the SCHZ group achieved higher correlation values across simulations, suggesting that the model reproduces functional connectivity more accurately in patients than in controls under the hemispheric-specific coupling scheme.\\

\begin{figure}[H]
    \centering
    \includegraphics[width=1.0\textwidth]{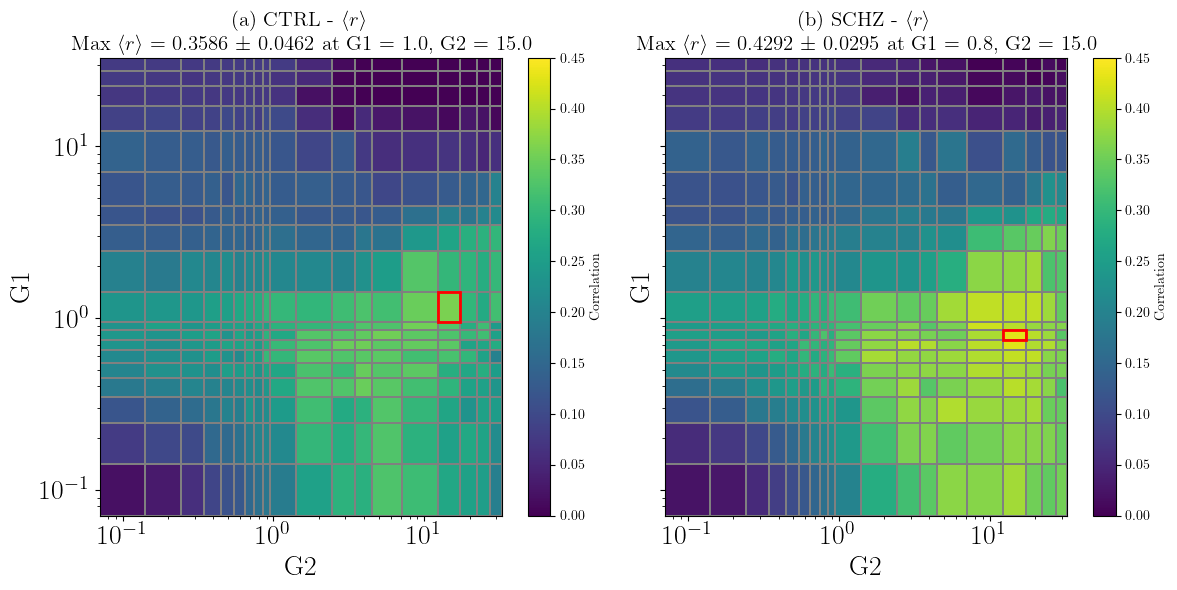}
    \caption{
    Parameter sweep across intra-hemispheric and inter-hemispheric coupling strengths \( G_1 \) and \( G_2 \) in the Wilson--Cowan model.  
    \textbf{(a,b)} $\braket{r}$ between simulated and empirical FC for the CTRL and SCHZ groups, respectively.  
    The red box in panels (a) and (b) marks the parameter combination that yielded the highest average correlation for each group. 
    }
    \label{fig:heatmap_correlation_rmse_G1_G2}
\end{figure}

Fig~(\ref{fig:heatmap_correlation_rmse_G1_G2}) summarizes the results of a parameter sweep over a wide range of \( G_1 \) and \( G_2 \) values. Figs~(\ref{fig:heatmap_correlation_rmse_G1_G2}a,b) show $\braket{r}$ between simulated and empirical FC for the CTRL and SCHZ groups, respectively. We observe a substantial improvement in performance compared to the uniform coupling case. The optimal configuration for the CTRL group was found at \( G_1 = 1.0 \), \( G_2 = 15.0 \), yielding a $\braket{r} = 0.36 \pm 0.05$ in Fig~(\ref{fig:heatmap_correlation_rmse_G1_G2}a). The SCHZ group achieved an even higher maximum $\braket{r} = 0.43 \pm 0.03$ at \( G_1 = 0.8 \), \( G_2 = 15.0 \) in Fig~(\ref{fig:heatmap_correlation_rmse_G1_G2}b). We observe that, for the CTRL group, the best-performing simulations are located in a more central region of the \( (G_1, G_2) \) parameter space compared to the SCHZ group. In the SCHZ case, the region of highest correlation is shifted toward higher \( G_2 \) values relative to \( G_1 \), indicating a stronger reliance on inter-hemispheric coupling. Moreover, the maximum correlation achieved in the SCHZ group is higher than that of the CTRL group. When comparing the mean correlation peaks, both groups share a same optimal value for \( G_2 \), but differ in their optimal \( G_1 \).  Specifically, the CTRL group requires stronger intra-hemispheric coupling to better replicate empirical FC patterns, whereas the SCHZ group achieves optimal performance with lower \( G_1 \). Although not shown, we also computed the RMSE between empirical and simulated connectivity patterns across the same parameter grid. Consistent with the correlation analysis, we found that configurations yielding higher correlations also exhibited lower RMSE values, indicating convergence in both topological correlation and absolute distance. 

While the model captures certain qualitative tendencies, such as the correlation between empirical and simulated FC, it does not yet reproduce network metrics comparable to those observed in the empirical data. This suggests that additional refinements are needed to improve the model's ability to replicate the topological organization of functional connectivity.

\section{Conclusions}

In this study, we implemented the Wilson--Cowan model and introduced a hemispheric-specific coupling scheme. We examined both groups and compared structural connectivity network features. In structural connectivity, we found results consistent with previous studies reporting decreased anatomical connectivity strength~\cite{griffa2013, guo2023}, redistribution of central roles across network regions~\cite{guo2023}, and topological reorganization in schizophrenia~\cite{griffa2015}. To further explore the mechanistic basis of these alterations, we evaluated the capacity of a biologically grounded neural mass model based on Wilson--Cowan dynamics~\cite{wilson1972, wilson1973, abeysuriya2018}, in order to reproduce resting-state functional connectivity from empirical cortical structural connectomes in matched control and schizophrenia groups. In our implementation, we differentiated intra-hemispheric and inter-hemispheric anatomical projections via a hemispheric-specific coupling scheme. This modified version systematically outperformed the classic global coupling approach, yielding higher Pearson correlations between simulated and empirical FC across both groups. The improvement was especially pronounced in the schizophrenia group in both models. This may reflect a stronger constraint of functional dynamics by the underlying structural scaffold in schizophrenia, where reduced global integration and increased modularity limit the repertoire of functional configurations. In such conditions, anatomically grounded models may more accurately capture the structure–function mapping~\cite{deco2012,deco2013}. Interestingly, the optimal parameter configuration differed between groups. The SCHZ group achieved a higher maximum correlation of \(\langle r \rangle = 0.43 \pm 0.03\) at \(G_1 = 0.8\), \(G_2 = 15.0\), whereas the CTRL group reached \(\langle r \rangle = 0.36 \pm 0.05\) at \(G_1 = 1.0\), \(G_2 = 15.0\). Although both groups showed optimal performance in regimes where inter-hemispheric coupling \( G_2 \) exceeded intra-hemispheric coupling \( G_1 \), the schizophrenia group required a lower \( G_1 \) to reproduce empirical FC. This may suggest a diminished role of intra-hemispheric integration in shaping functional dynamics in schizophrenia, potentially reflecting degraded or less functionally effective intra-hemispheric anatomical pathways~\cite{guo2023}. Additionally, the SCHZ model displayed good performance across a broader range of \( G_2 \) values, indicating a degree of flexibility or compensatory reliance on inter-hemispheric interactions. Together, these findings suggest distinct structure–function coupling regimes between groups, with schizophrenia exhibiting both reduced intra-hemispheric integration and a more tolerant configuration space for inter-hemispheric coupling. Overall, these findings highlight the value of integrating structural connectivity data with biologically inspired models to gain mechanistic insights into the altered brain dynamics observed in neuropsychiatric disorders.

\section*{Data Availability and Reproducibility}

All simulations were implemented in Python 3.11. The implementation relies on the following open-source libraries: \texttt{NumPy 1.26}, \texttt{SciPy 1.13}, \texttt{pandas 2.2.1}, \texttt{matplotlib 3.8.3}, and \texttt{Numba 0.60} for just-in-time compilation of performance-critical functions. The Wilson--Cowan model and analysis scripts are structured for modularity and extensibility. The entire simulation pipeline can be reproduced by running the provided main scripts (\texttt{run\_global.py} and \texttt{run\_hemispheric.py}). A \texttt{requirements.txt} file is provided to facilitate environment replication. The simulation outputs (e.g., empirical and simulated functional connectivity matrices, node activity signals, and computed metrics) are stored in a structured directory under \texttt{results/}, which is automatically generated by the code. These outputs include all data used to produce the figures presented in this work. The data that support the findings of this study are openly available in \textcolor{blue}{\texttt{\href{https://github.com/ramirop2021/Wilson-Cowan-Hemispheric-Coupling}{GitHub}}}.

\bibliographystyle{unsrt}

\end{document}